\documentclass[a4paper,fleqn,usenatbib]{mnras}

\usepackage{newtxtext,newtxmath}

\usepackage[T1]{fontenc}
\usepackage{ae,aecompl}
\usepackage{natbib,times}
\usepackage{array}

\usepackage{graphicx}	

\title[The masses of RSG progenitors to SNe]{The Initial Masses of the Red Supergiant Progenitors to Type-II Supernovae}

\author[B. Davies \& E.R. Beasor]{
Ben Davies,$^{1}$\thanks{E-mail: b.davies@ljmu.ac.uk}
and Emma R.\ Beasor$^{1}$
\\
$^{1}$Astrophysics Research Institute, Liverpool John Moores 
University, Liverpool Science Park ic2, \\ 146 Brownlow Hill, Liverpool, L3 5RF, UK\\
}

\date{Accepted XXX. Received YYY; in original form ZZZ}

\pubyear{2017}

\begin{document}
\label{firstpage}
\pagerange{\pageref{firstpage}--\pageref{lastpage}}
\maketitle

\def\ga{\mathrel{\hbox{\rlap{\hbox{\lower4pt\hbox{$\sim$}}}\hbox{$>$}}}}
\def\la{\mathrel{\hbox{\rlap{\hbox{\lower4pt\hbox{$\sim$}}}\hbox{$<$}}}}
\def\msunyr{$M$ \mbox{$_{\normalsize\odot}$}\rm{yr}$^{-1}$}
\def\msun{$M$\mbox{$_{\normalsize\odot}$}}
\def\zsun{$Z$\mbox{$_{\normalsize\odot}$}}
\def\rsun{$R$\mbox{$_{\normalsize\odot}$}}
\def\minit{$M_{\rm init}$}
\def\lsun{$L$\mbox{$_{\normalsize\odot}$}}
\def\mdot{$\dot{M}$}
\def\mdotdj{$\dot{M}_{\rm dJ}$}
\def\lbol{$L$\mbox{$_{\rm bol}$}}
\def\kms{\,km~s$^{-1}$}
\def\EW{$W_{\lambda}$}
\def\arcsec{$^{\prime \prime}$}
\def\arcmin{$^{\prime}$}
\def\teff{$T_{\rm eff}$}
\def\Teff{$T_{\rm eff}$}
\def\logg{$\log g$}
\def\logz{$\log Z$}
\def\vdisp{$v_{\rm disp}$}
\def\um{$\mu$m}
\def\chisq{$\chi^{2}$}
\def\AV{$A_{V}$}
\def\hminus{H$^{-}$}
\def\Hminus{H$^{-}$}
\def\ebmv{$E(B-V)$}
\def\mdyn{$M_{\rm dyn}$}
\def\mphot{$M_{\rm phot}$}
\def\cnterm{[C/N]$_{\rm term}$}
\newcommand{\fig}[1]{Fig.\ \ref{#1}}
\newcommand{\Fig}[1]{Figure \ref{#1}}
\newcommand{\refc}[1]{{#1}}

\def\mlo{$M_{\rm lo}$}
\def\mhi{$M_{\rm hi}$}
\def\BCV{{\it BC}$_V$}
\def\BCR{{\it BC}$_R$}
\def\BCI{{\it BC}$_I$}
\def\BCK{{\it BC}$_K$}
\def\BCl{{\it BC}$_\lambda$}
\def\lbol{$L_{\rm bol}$}
\def\logl{$\log(L/L_{\odot}$}
\def\lfin{$L_{\rm fin}$}

\begin{abstract}
There are a growing number of nearby SNe for which the progenitor star is detected in archival pre-explosion imaging. From these images it is possible to measure the progenitor's brightness a few years before explosion, and ultimately estimate its initial mass. Previous work has shown that II-P and II-L supernovae (SNe) have Red Supergiant (RSG) progenitors, and that the range of initial masses for these progenitors seems to be limited to $\la$17\msun. This is in contrast with the cutoff of 25-30\msun\ predicted by evolutionary models, a result which is termed the 'Red Supergiant Problem'. Here we investigate one particular source of systematic error present in converting pre-explosion photometry into an initial mass, that of the bolometric correction (BC) used to convert a single-band flux into a bolometric luminosity. We show, using star clusters, that RSGs evolve to later spectral types as they approach SN, which in turn causes the BC to become larger. Failure to account for this results in a systematic underestimate of a star's luminosity, and hence its initial mass. Using our empirically motivated BCs we reappraise the II-P and II-L SNe that have their progenitors detected in pre-explosion imaging. Fitting an initial mass function to these updated masses results in an increased upper mass cutoff of \mhi=$19.0^{+2.5}_{-1.3}$\msun, with a 95\% upper confidence limit of $<$27\msun. Accounting for finite sample size effects and systematic uncertainties in the mass-luminosity relationship raises the cutoff to \mhi=25\msun\ ($<$33\msun, 95\% confidence). We therefore conclude that there is currently no strong evidence for `missing' high mass progenitors to core-collapse SNe. 
\end{abstract}



\section{Introduction} \vspace{1cm}
The direct identification of the progenitor star to core-collapse supernovae (SNe) provides a critical test of the predictions of stellar evolution. Specifically, various evolutionary codes have predicted that single stars with masses between $\sim$8 and 25-30\msun\ will end their lives in the Red Supergiant (RSG) phase, the progenitor's H-rich envelope will cause the SN to be classified as a II-plateau or II-long decline (II-P/II-L), and the explosion will leave behind a neutron star remnant \citep[e.g.][]{Ekstrom12,Heger03}. 

These predictions can be tested using SNe in nearby galaxies, which often have archival deep imaging data obtained for other science goals. Following pioneering work by \citet{Smartt01SN} and \citet{Smartt04}, accurate image alignment has been used to identify the progenitor star for 14 Type-IIP/L SNe to date, and in a few cases late-time imaging has been subtracted from the pre-explosion data to prove the progenitor has disappeared, and to provide accurate pre-explosion photometry \citep[e.g.][]{Maund09}. For a review of this work, see \citet[][ hereafter S15]{Smartt15}.

Once the progenitor has been identified, its pre-explosion brightness can be used to estimate its initial mass. With a sample of progenitors studied in this way, it is possible to compare the distribution of progenitor masses with that expected from a standard initial mass function (IMF) and upper/lower limits to the mass range. Such analysis yielded a surprising result; specifically that the upper limit to the mass range was $\sim$17\msun, as opposed to the expected 25-30\msun\ \citep[][ hereafter S09; S15]{Smartt09}. This result was termed `the RSG problem', and has received much attention in the literature. \refc{Possible explanations include substantially increasing RSG mass-loss rates to lower the initial mass at which the H-rich envelope is lost prior to SN \citep{Ekstrom12,Georgy12}, or lowering the mass threshold at which black-holes may form at core-collapse (S09). The latter explanation would resonate across} many areas of astrophysics, such as the formation rate of black-holes, the cosmic SN rate, and the significance of SN feedback in galaxy evolution \citep[e.g.][]{Kochanek15,Horiuchi11,Crain15}. 

Turning the pre-explosion fluxes into an initial mass is a two-step process. Firstly, the flux must be turned into a bolometric luminosity using either model fitting to the spectral energy distribution (SED) \citep[e.g.][]{Maund14bk}, or if the pre-explosion data is limited to one or two passbands, by assuming a bolometric correction (BC). Secondly, the final luminosity \lfin\ is converted into an initial mass \minit\ via numerical predictions of the relation between \minit-\lfin\ from evolutionary models (for details on this process, see S09). 

In this paper we concentrate on the first step, that is the conversion of fluxes to bolometric luminosities. The first potential source of systematic error is the issue of extinction due to circumstellar material (CSM), which many RSGs are known to have, often at the level of $A_V \ga 1$mag \citep[e.g.][]{Massey05,dewit08,rsgteff}. If this is not accounted for, \lfin\ (and hence \minit) will be underestimated \citep{Walmswell-Eldridge12,Beasor-Davies16}. However, \citet{Kochanek12} have argued that this effect would be small, since photons scattered into the line of sight by circumstellar dust can almost entirely compensate for the direct flux lost, provided the CSM is not too optically thick. In S15, the progenitor mass distribution from S09 was re-analysed including an ad-hoc level of circumstellar extinction of $A_V=0.5$, and it was argued that the updated upper mass cutoff of $\sim$17\msun was still much lower than the predicted 25-30\msun. 

Another potential source of systematic error is the adopted BC. As stated above, for many SN progenitors there are pre-explosion detections in only one or two bands. With a single-band detection, one cannot know the progenitor's colour; while with a two-band detection one can only place an upper limit on the colour under the assumption of zero reddening. In these cases, the strategy is typically to assume that the progenitor was a RSG, and assume an average BC based on the average spectral type of RSGs. Typically, a spectral type of M0 $\pm$3 subtypes is assumed, which corresponds to bolometric correction in the $V$-band of \BCV=$-1.3\pm0.5$ (see e.g. S09, S15). By adopting such a Gaussian probability distribution for \BCV, the resulting initial mass estimate is exposed to a potential systematic error if the true BC has a larger (negative) value. 

In this paper we reassess the BCs of RSGs, in particular the evolution of BCs in various passbands as the stars approach SN. We do this by studying RSGs in resolved young star clusters. By studying stars in clusters, we can be confident that the stars all have approximately the same age, and so the stars currently in the RSG phase have the same initial mass to within a few tenths of a Solar mass. Since these RSGs will all be following roughly the same evolutionary track as they climb up the Hertzsprung-Russell (H-R) diagram, we can use their luminosity as a proxy for evolutionary state \citep[see e.g.\ Fig.\ 1 of][]{rsgc1paper}. That is, we expect the most luminous RSG in a cluster to be the star closest to SN. Therefore, by measuring the properties of the RSGs in a cluster as a function of luminosity, we can study how these properties change as the stars evolve. Finally, by studying RSGs in clusters where we can clearly detect the underlying main-sequence (MS) population, we can use the location of the MS in the H-R diagram to make an independent estimate of the foreground extinction. This is a crucial advantage if we wish to study the intrinsic colour evolution of the RSGs. 

This paper is organised as follows. In Sect.\ \ref{sec:data} we describe the clusters in our sample, the sources for the photometry of the RSGs studied here, our de-reddening procedure, and how we determine the BCs for each star. In Sect.\ \ref{sec:evol} we investigate the trends of spectral type and BC with luminosity, and hence evolutionary state. In Sect.\ \ref{sec:prog} we use our results to reassess the progenitor masses for all II-P and II-L SNe with pre-explosion photometry or upper limits, and ultimately revisit the mass distribution of SN progenitors and the upper mass cutoff. We conclude in Sect.\ \ref{sec:conc}. 

\begin{table*}
\caption{The extinction law for RSGs. The first column contains the name of the filter bandpass $\lambda$. The next three column blocks contain the extinction at a given filter $\lambda$ for an early stage (spectral type = M0) and late-stage (M7) RSG, for reddening values of \ebmv=0.322, 1.61 and 3.22 -- analogous to visual band extinctions of $A_V = $1, 5 and 10 for a \citet{cardelli89} reddening law with $R_V=3.1$. See text for further details. }
\begin{center}
\setlength{\extrarowheight}{6pt}
\begin{tabular}{l|cc|cc|cc}
\hline \hline
 & \multicolumn{2}{c|}{\ebmv=0.322} & \multicolumn{2}{c|}{\ebmv=1.61} & \multicolumn{2}{c}{\ebmv=3.22} \\ 
 & \multicolumn{2}{c|}{$A_\lambda / A_V$} & \multicolumn{2}{c|}{$A_\lambda / A_V$} & \multicolumn{2}{c}{$A_\lambda / A_V$} \\ 
$\lambda$ & (M0) & (M7) & (M0) & (M7) & (M0) & (M7) \\
\hline
 JOHNSON U & 1.285 & 1.050 & 1.402 & 1.285 & 1.449 & 1.374 \\
  JOHNSON B & 1.215 & 1.199 & 1.194 & 1.179 & 1.171 & 1.158 \\
  JOHNSON V & 0.972 & 0.969 & 0.961 & 0.957 & 0.945 & 0.941 \\
  JOHNSON R & 0.720 & 0.698 & 0.697 & 0.677 & 0.671 & 0.652 \\
  JOHNSON I & 0.504 & 0.500 & 0.491 & 0.487 & 0.477 & 0.474 \\
WFPC2 F814W & 0.544 & 0.536 & 0.531 & 0.524 & 0.517 & 0.511 \\
    2MASS J & 0.338 & 0.426 & 0.293 & 0.273 & 0.294 & 0.266 \\
    2MASS H & 0.182 & 0.181 & 0.181 & 0.181 & 0.181 & 0.180 \\
    2MASS K & 0.119 & 0.119 & 0.118 & 0.118 & 0.118 & 0.118 \\
\hline
\end{tabular}
\end{center}
\label{tab:redden}
\end{table*}%

\section{Observations and data} \label{sec:data}

In selecting clusters for this work, we have the following requirements. Firstly, the clusters must be young enough to contain RSGs ($\la$25Myr), whilst being massive enough to have a well-populated upper IMF ($\ga 10^4$\msun). The clusters must also be nearby enough such that we can resolve the individual stars, and that the line-of-sight extinction is low enough for us to be able to detect the stars in the optical. The latter requirement is necessary as we are aiming to derive empirical bolometric luminosities, so we need to be able to detect the stars at any wavelength at which they emit non-negligible flux. 

From the above requirements we have selected four clusters: the galactic clusters $\chi$~Per, NGC7419, and RSGC2, and NGC2100 in the Large Magellanic Cloud. Below we discuss each of the clusters in more detail.

\subsection{$\chi$~Per}
This cluster is part of the larger $h+\chi$~Per complex, all of which appears to be coeval to within the errors on the ages of the constituent components \citep{Currie10}. This complex is itself part of the larger Per~OB1 complex. To ensure that we are only selecting coeval stars, we have restricted our sample to those RSGs within 6\arcmin\ of the centre of $\chi$~Per, which is approximately the distance to the edge of the $h+\chi$~Per complex. This leaves us with a sample of 8 stars. Spectral types are taken from the SIMBAD database. The optical extinction towards $\chi$~Per was determined by \citet{Currie10} from fits to the location of the main-sequence stars, and was found to be $A_V$=1.66 mag. 

The majority of the optical photometry is taken from \citet{Johnson66}. The stars in this cluster have more contemporary photometry available in the literature \citep[e.g.][]{Kharchenko09,Pickles10}, enabling us to assess the reliability of the original Johnson photometry\footnote{The \citet{Pickles10} data is synthetic photometry on the Landolt system constructed from various multiband photometry in the literature.}. The photometry at {\it UBV} is consistent to within $\sim$0.1mags, which seems reasonable considering some of these stars are likely to be photometrically variable. However, the $I$ band photometry from different authors disagrees at a level which cannot be explained by stellar variability. Specifically, the Pickles synthetic Cousins $I \equiv I_C$ band photometry and Johnson $I_J$ photometry are offset from one another by around half a magnitude for all stars, even after colour-corrections to put them in the same photometric system (see Appendix). Furthermore, neither set of $I$ photometry produces SEDs for the stars which look typical of RSGs. As a compromise, for the $I$ band photometry in this cluster we have taken an average of the Johnson and Pickles photometry, with the error defined to be half the difference between the two.

\subsection{NGC~7419} 
This cluster is relatively compact, containing five RSGs, including the relatively extreme RSG MY~Cep. Spectral types are again taken from the SIMBAD database. As with $\chi$~Per, main-sequence fitting reveals an optical extinction of $A_V=5.2$ \citep{Joshi08,Marco13}. The optical {\it UBVRI} photometry for these stars is taken from \citet{Joshi08}, and is complemented with near- and mid-IR photometry from 2MASS \citep{skrutskie06} and MSX \citep{egan03}. The $I$-band photometry has been converted from the Cousins to the Johnson system (see Appendix).

\subsection{RSGC2}
This cluster is much more heavily extinguished than the rest in our sample, so much so that it has yet to be detected at wavelengths shorter than $V$. A population of blue stars was detected in the optical by \citet{Nakaya01}, the colours of which imply a $V$-band extinction of $A_V$=11.4$\pm$0.6. This is comparable to the $K$-band extinction $A_K \sim 1$ found by \citet{Froebrich-Scholz13} who detected the main-sequence in the near-IR. However, fundamental uncertainties in the nature of the extinction law make correcting for this extinction highly problematic, as we will discuss further in Sect. \ref{sec:bc}.

The cluster has 26 RSGs \citep{rsgc2paper}, but only a subsample of these have known optical spectral types \citep{Negueruela12,Negueruela13}. These stars have high quality near and mid-IR photometry from 2MASS, {\it Spitzer/IRAC} \citep{Benjamin03} and MSX. Optical photometry in the $V$, $R$ and $I$ bands comes from \citet{Nakaya01}, which according to the authors is photometrically similar to the Johnson system.

\subsection{NGC~2100}
This cluster, located in the Large Magellanic Cloud (LMC), has 19 RSGs. Optical photometry in the {\it UBVR} bands is taken from \citet{Massey02}. This is complemented with $I$-band photometry from DENIS \citep{Cioni00}, near-IR photometry from 2MASS, and mid-IR photometry from WISE \citep{WISE}. Spectral types are available only for a small subsample of stars in this cluster, obtained from I.\ Negueruela (priv.\ comm). For RSGs, the DENIS-$I$ bandpass has a colour correction to the Johnson system which for RSGs is indistinguishable from that of Cousins-$I$ (see Appendix).

\subsection{Dereddening the photometry} \label{sec:extinct}
For the clusters $\chi$~Per and NGC2100, the low foreground extinction requires only a minor correction, and so the uncertainty introduced by the dereddening process is minor compared to other sources of error such as the errors on the photometry. However, for NGC7419, and especially for RSGC2, uncertainties in the foreground extinction and the extinction law are the dominant source of error. 

In making these corrections, the first factor we need to appreciate is that the commonly-used extinction laws used to convert a reddening \ebmv\ to extinction at a particular wavelength $A_\lambda$ \citep[e.g.][]{cardelli89} were derived using OB stars, which have very different spectral energy distributions (SEDs) to RSGs. This means that the tabulated $A_\lambda / A_V$ values \citep[e.g.][]{R-L85} no longer hold, as the effective wavelength for the filter $\lambda$ can be substantially redder when applied to a RSG. Indeed, there is a second-order effect, in that the higher the total extinction the redder the effective wavelength of the filter. This has been previously noted in the literature \citep{Johnson-Mendoza66,Lee70,Elias85,Nakaya01}. Given our previously mentioned concerns about some of the literature photometry for bright RSGs in the $I$-band, plus the fact that filter bandpasses frequently used to detect SN progenitors can be very different from those of the Johnson system, we have decided to revisit this issue with contemporary data. 

To determine the appropriate reddening corrections, we began with the observed SEDs of a sample of RSGs first presented in \citet{rsgteff}. The sample covers RSGs of spectral types K0 to M2.5, with continuous wavelength coverage between 0.35\um\ and 2.5\um. We derived synthetic photometry for each star for a range of filters on the magnitude scale, $m_\lambda$. We then reddened the spectra according to \citet{cardelli89}, using a canonical value of $R_V=3.1$ and a fixed value of \ebmv, and again determined synthetic photometry for the same set of filters, $m^\prime_\lambda$. The extinction at a given filter $A_\lambda$ for a fixed \ebmv\ was then determined to be $A_\lambda = m^\prime_\lambda - m_\lambda$.  

In Table \ref{tab:redden} we have tabulated values of $A_\lambda$ for various filters, including the F814W filter that frequently provides the detections for II-P progenitors in pre-explosion imaging. We have provided the correction for spectral type M0, as well as the correction for an M7 star if we take our data and linearly extrapolate the trend with spectral type. We have also provided the corrections for three levels of extinction, analogous to values of $A_V$ of 1, 5, and 10 mags.

\begin{table}
\caption{Stars used in this study, along with their spectral types and the bolometric luminosities derived in this work. The errors on \lbol\ do not include the errors in the distance to the host cluster. The identifications come from \citet{Beasor-Davies16}, \citet{Beauchamp94} and \citet{rsgc2paper}. }
\begin{center}
\setlength{\extrarowheight}{6pt}
\begin{tabular}{lcc}
\hline \hline
ID & SpT & $\log(L_{\rm bol}/L_\odot)$ \\
\hline
{\footnotesize \it NGC 2100} \\
BD       1 &   M6 & 4.87$^{+0.01}_{-0.01}$ \\        
BD       3 &   M0 & 4.84$^{+0.01}_{-0.01}$ \\        
BD       7 &   M2 & 4.73$^{+0.01}_{-0.01}$ \smallskip\\        
\hline
{\footnotesize \it NGC 7419} \\
BMD 950 & M7.5 & 5.25$^{+0.04}_{-0.04}$ \\        
BMD 435 & M1.5 & 4.62$^{+0.02}_{-0.01}$ \\        
BMD 696 & M1.5 & 4.62$^{+0.04}_{-0.03}$ \\        
BMD 139 &   M1 & 4.53$^{+0.04}_{-0.04}$ \\        
BMD 921 &   M0 & 4.47$^{+0.05}_{-0.05}$ \smallskip\\        
\hline
{\footnotesize \it $\chi$ Per} \\
 SU Per  & M3.5 & 4.99$^{+0.09}_{-0.06}$ \\        
 RS Per  &   M4 & 4.92$^{+0.06}_{-0.04}$ \\        
 AD Per  & M2.5 & 4.80$^{+0.08}_{-0.05}$ \\        
V441 Per &   M2 & 4.75$^{+0.10}_{-0.06}$ \\        
 BU Per  &   M4 & 4.67$^{+0.07}_{-0.05}$ \\        
 FZ Per  &   M1 & 4.64$^{+0.06}_{-0.05}$ \\        
V439 Per &   M0 & 4.53$^{+0.06}_{-0.05}$ \\        
V403 Per &   M0 & 4.41$^{+0.06}_{-0.05}$ \smallskip\\       
\hline
{\footnotesize \it RSGC2} \\
 D2 & M7.5 & 5.12$^{+0.14}_{-0.20}$ \\        
 D6 & M3.5 & 5.04$^{+0.01}_{-0.37}$ \\        
 D3 &   M5 & 4.94$^{+0.14}_{-0.26}$ \\        
D18 & M0.5 & 4.66$^{+0.19}_{-0.01}$ \\        
D16 &   M2 & 4.56$^{+0.11}_{-0.30}$ \\        
D20 & M1.5 & 4.56$^{+0.15}_{-0.35}$ \\        
D19 &   M1 & 4.45$^{+0.26}_{-0.25}$ \smallskip \\        
\hline
\end{tabular}
\end{center}
\label{default}
\end{table}%

\subsection{Determination of \lbol\ and \BCl}
With all the photometry and spectrophotometry for each star collated, we next computed bolometric luminosities. We interpolated between photometry points using a cubic spline in log space, and integrated between 0.3 and 24\um. For the stars in RSGC1 which had photometry missing in the $U$ and $B$ bands, we extrapolated the SED from the $V$-band to the $U$-band using a black-body curve that had been fitted to the available photometry. This extrapolated region of the SED accounted for around 0.1dex of the total flux of the star. Though the SEDs of RSGs are poorly matched by black-bodies, tests using the other stars in our sample revealed that the error introduced by this procedure was small ($\la \pm$0.03dex) compared to the other sources of error in these stars. 

The bolometric corrections at each band \BCl\ were determined from, 

\begin{equation}
{\rm BC}_\lambda = m_{\rm bol} - m_\lambda
\end{equation} 

\noindent where $m_\lambda$ is the apparent magnitude at wavelength $\lambda$, corrected for foreground extinction, and $m_{\rm bol}$ is the bolometric apparent magnitude of the star as determined above. 

We note that we have {\it not} corrected for any circumstellar extinction here. RSGs are well known to have large amounts of circumstellar dust produced in their winds, and that the mass of this dust increases as the star approaches SN \citep{Beasor-Davies16}. Here we are assuming that any flux lost in the optical and near-IR bands due to absorption or scattering by the dust is re-radiated in the mid-IR. Hence, our \lbol\ measurements are unaffected by circumstellar extinction, and the \BCl\ values we derive for each star in our sample should include any correction for circumstellar dust. If all RSGs at a certain spectral type are affected by circumstellar extinction in a similar way, this should reveal itself in the average \BCl\ for that spectral type.

\section{Trends between evolutionary state observed properties} \label{sec:evol}

\subsection{Evolution of spectral type}

\begin{figure}
\centering
\includegraphics[width=8.5cm]{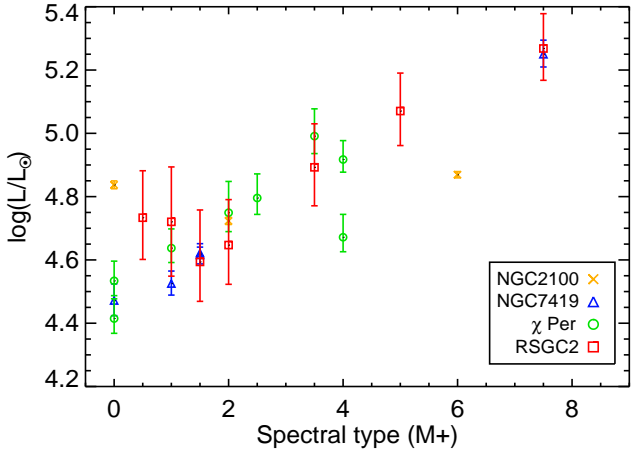}
\caption{The correlation of spectral type with luminosity in four resolved star clusters. Since the most luminous RSGs in a star cluster are the most evolved, we conclude that the RSGs with the latest spectral type are the closest to SN.}
\label{fig:lum-spt}
\end{figure}

The first trend we remark upon is that of evolutionary state (i.e. luminosity) with spectral type, which is shown in \fig{fig:lum-spt}. We see a clear trend of the most luminous (and hence most evolved) RSGs having later spectral types. Such a correlation has already been noted by \citet{Negueruela13}. Three of the four clusters studied here have stars with spectral types M6 or later, with the latest spectral type in $\chi$~Per being M4. However, the stars W~Per and S~Per are part of the larger Per~OB1 complex, and so may conceivably have the same age, and both are known to have shown spectral types as late as M5 and M7 respectively. Together with the data plotted in \fig{fig:lum-spt}, this leads us to conclude that the spectral type of RSGs evolves to later types as the star approaches the end of its life, and that the spectral type at SN will be typically around M5-M7.

The above conclusion is consistent with that found for SN progenitors which have either pre-explosion detections in multiple bands or stringent upper limits. The progenitor of SN2009md had colours consistent with being an M4 \citep{Fraser11}, while SN2004et was redder than an M4 \citep{Crockett11}. SN2003gd and SN2013ej had colours consistent with M2 supergiants \citep{vd03,Smartt04,Fraser14}, while the recently revised foreground extinction towards SN2008bk also implies a late-type progenitor \citep[this work, ][]{Maund17}. Other SNe II-P which were thought to have yellow progenitors have since been called into question \citep{Maund15}. From SN progenitor evidence alone, one would conclude that Type-IIP/L SNe should typically have RSG progenitors with spectral types of at least M2, and in most cases significantly later. 

\begin{figure*}
\centering
\includegraphics[width=8.5cm]{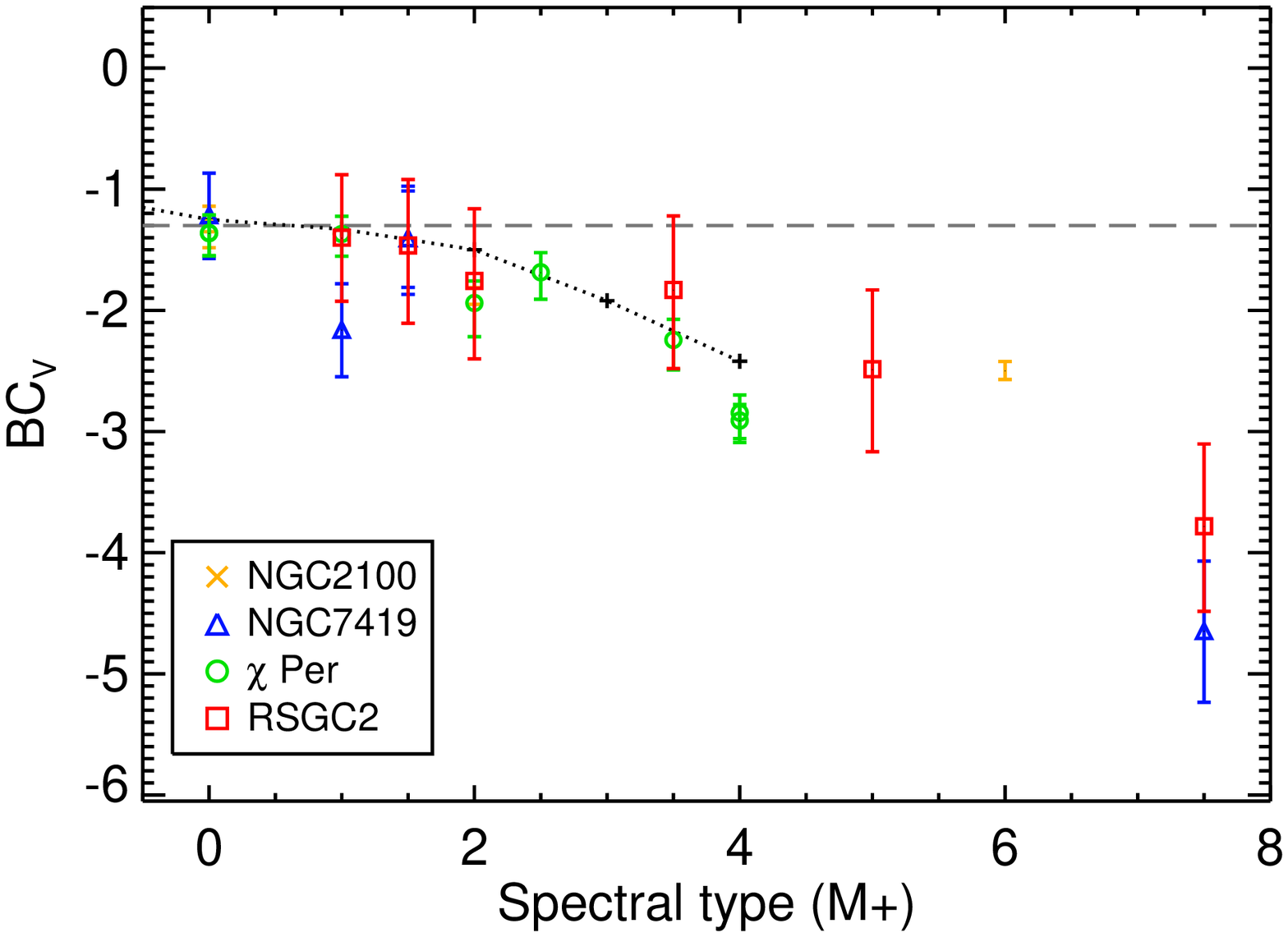}
\includegraphics[width=8.5cm]{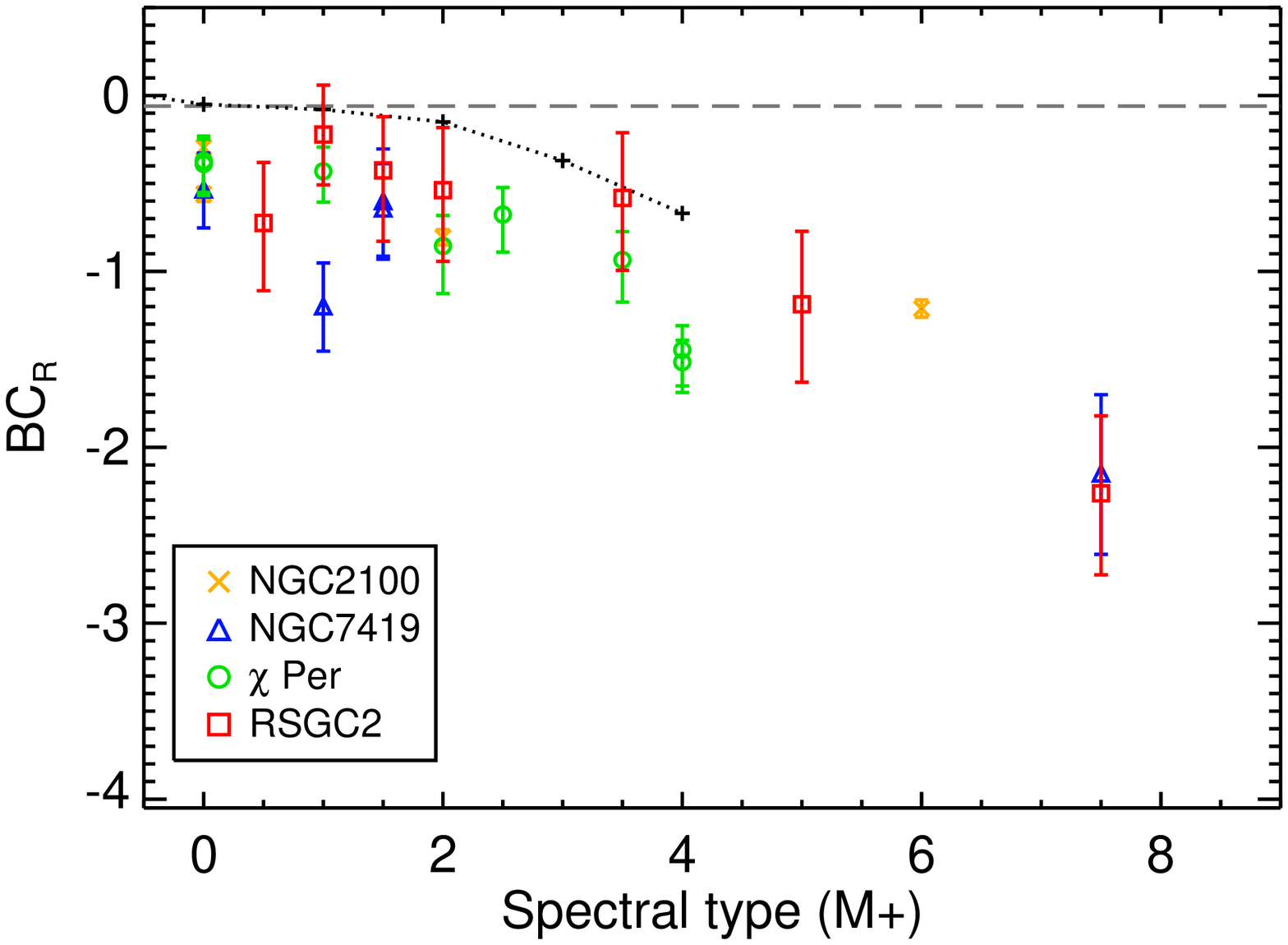}
\includegraphics[width=8.5cm]{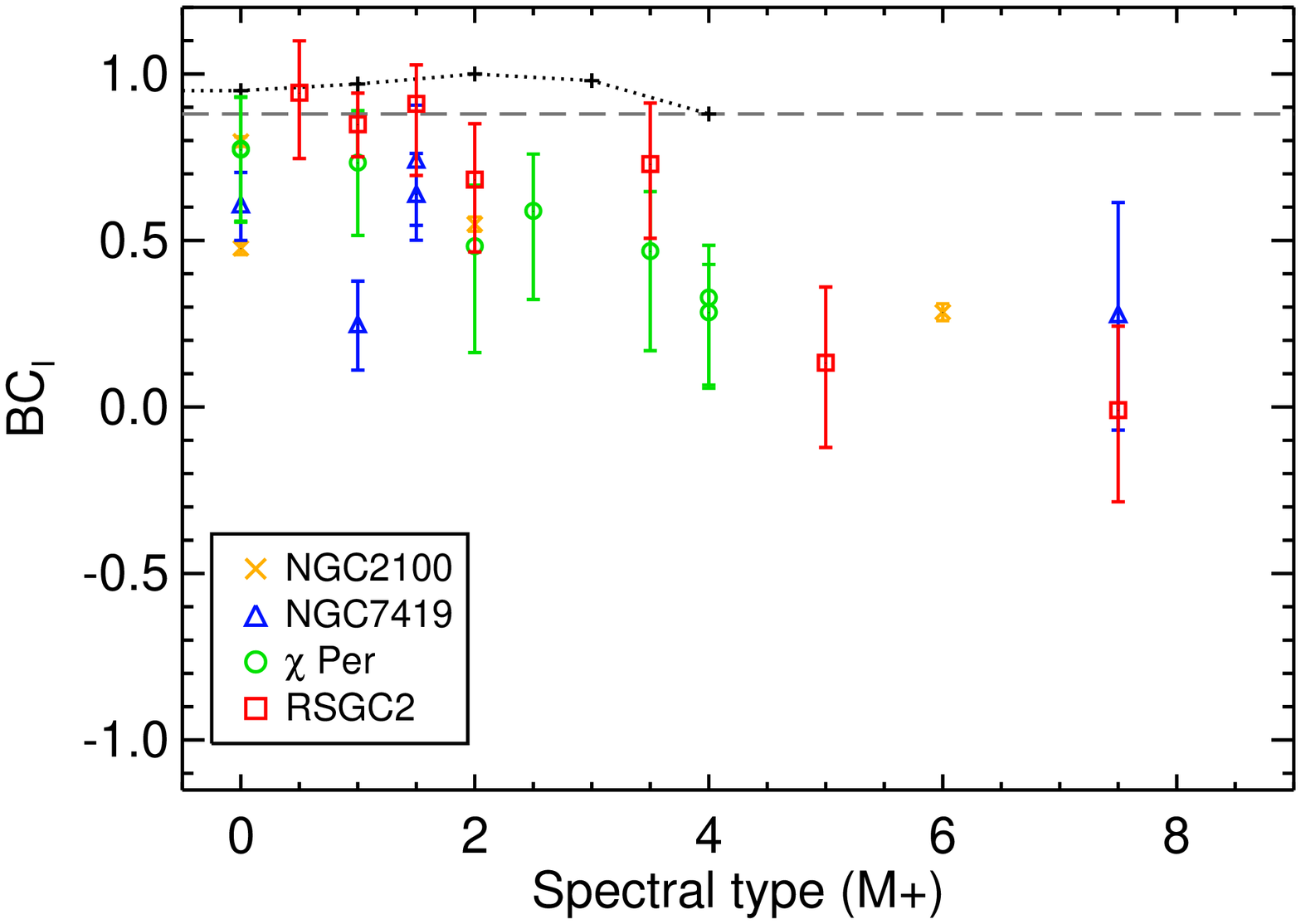}
\includegraphics[width=8.5cm]{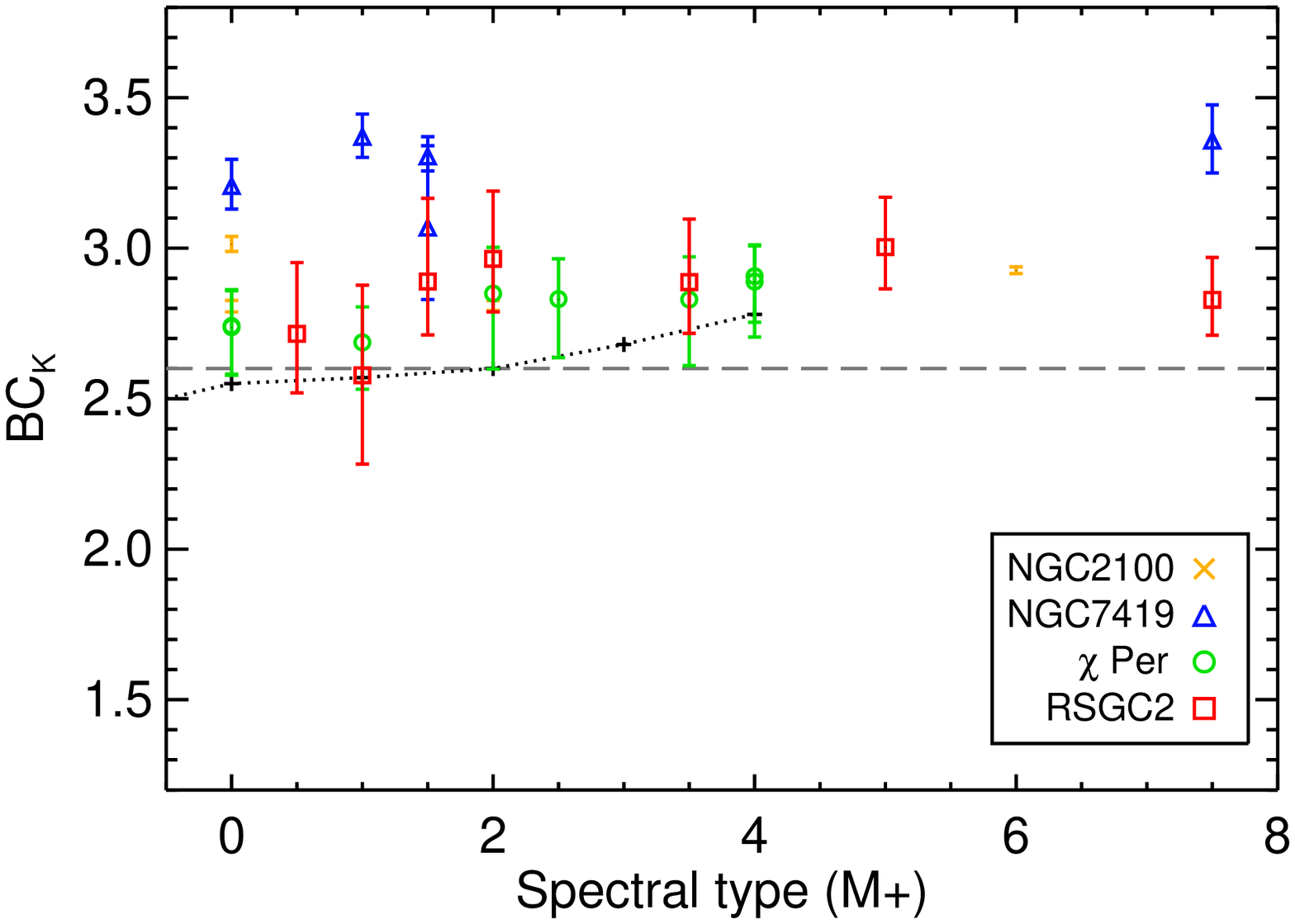}
\caption{Bolometric corrections (BCs) for each of the stars in each cluster. The panels show the BCs in $V$, $R$, $I$ and $K$-bands (top-left, top-right, bottom-left, bottom-right respectively). The grey dashed lines shows the BCs adopted by S09 and S15 in each band. The dotted line with `+' symbols shows the BCs measured by \citet{Elias85}, which only go as far as M4. }
\label{fig:spt-bc}
\end{figure*}

\subsection{Evolution of RSG bolometric corrections} \label{sec:bc}
The bolometric corrections in the $V$, $R$, $I$ and 2MASS $K_s$ bands as a function of spectral type for the stars in our sample are shown in \fig{fig:spt-bc}. In the top left panel where we show \BCV, we see that for a spectral type of M0 the \BCV\ assumed by S09 and S15 are consistent with our early M-type stars. Our results are also consistent with the \BCV\ of \citet{Elias85}, illustrated by the dotted line. However, the plot shows that for spectral types later than those studied by Elias et al, \BCV\ drops very steeply. The average \BCV\ for the two M7.5 stars in our sample is -4; which is 2.7mags below that adopted by S09/S15. This would mean that for any SN progenitor with a single pre-explosion detection in the $V$-band, the luminosity would be underestimated by over a factor of 10. 

Fortunately, pre-explosion detections are more frequently in the $I$-band, and occasionally in the $R$-band (or their {\it HST} analogues, F814W and F606W respectively). For these bands, the evolution of BC with spectral type is less aggressive. Nevertheless, from our data the difference in BC between an M0 star and an M7.5 seems to be of order -2 and -1 mags in $R$ and $I$ respectively. In the $K$-band, the trend with spectral type is weak but seems to go in the opposite direction. This means that for a pre-explosion detection at $K$, the S09/S15 luminosity estimate would be slightly too high, though only by $\sim$0.15dex.  

On each panel of \fig{fig:spt-bc} we also show the BCs of \citet{Elias85}. In the $V$ and $K$ bands, we find results that are consistent with this work to within the errors. However, in the $R$-band, and in particular the $I$ band, the systematic differences are large. As noted earlier, the $I$-band photometry used by Elias et al. came from \citet{Johnson66}.  When we compared this photometry to that at other wavelengths for the same stars, the $I$-band seemed to be anomalously bright by around half a magnitude. We note also that the intrinsic colours and BCs derived by \citet{Fraser11} using MARCS model atmospheres with effective temperatures between 3400-4000K are comparable to our results, with the caveat that MARCS models notoriously struggle to fit the TiO bands that dominate the $V$ and $R$ bands of RSGs \citep{rsgteff}. We therefore suggest that a problem with the photometric calibration of the $I$ band in \citet{Johnson66} is the cause of the discrepancy between our results and those of Elias et al. 

To determine the most appropriate \BCl\ for RSGs late in their evolution, we take an error-weighted average of the latest star in all clusters: BD-1 in NGC~2100, BMD~950 ($\equiv$MY~Cep) in NGC~7419, and RS~Per and BU~Per in $\chi$~Per (both M4). We have decided against including RSGC2-D2 in this averaging procedure as the BCs in this cluster are heavily dependent on not just the high (and uncertain) foreground extinction, but also the slope of the extinction law. In the panels of \fig{fig:spt-bc}, we have shown the results for RSGC2 using a value of $R_V=4.5$, rather than the usual $R_V=3.1$. Such a value of $R_V$ is not uncommon for regions in the inner Galaxy \citep[e.g.][]{Zasowski09}. Using the standard  $R_V=3.1$ would cause the RSGC2 points in move downward in \fig{fig:spt-bc}. In the case of the $I$-band (bottom left panel) the points would move down by about a magnitude, and so would be highly discrepant with respect to the data from the other clusters. For the purposes of this paper, we show the RSGC2 points only to illustrate that by making justifiable assumptions about the extinction law in that line of sight we can make these stars agree with the rest. We exclude the RSGC2 stars when computing average \BCl\ values.

\begin{table}
\caption{Average bolometric corrections for late-type RSGs, as determined from the star with the latest spectral types (M4-M7.5) in the clusters $\chi$~Per, NGC~7419 and NGC~2100. See text for details.}
\begin{center}
\begin{tabular}{lc}
\hline \hline
$\lambda$ & \BCl \\
\hline
$V_J$     & -3.88 $\pm$ 0.82 \\
$R_J$     & -1.83 $\pm$ 0.38 \\
$I_J$       & 0.25$\pm$ 0.15 \\
$I_C$       & -0.32$\pm$ 0.15 \\
$F814W$  & 0.00$\pm$ 0.15 \\
${\it Gunn~i^\prime}$ & -0.49$\pm$ 0.15 \\
$K_S$  & 3.00 $\pm$ 0.18 \\
\hline
\end{tabular}
\end{center}
\label{tab:bcs}
\end{table}%

In Table \ref{tab:bcs} we compile the average BCs for late-type RSGs for several band-passes. This list of filter bands includes several definitions of the $I$-band, where the majority of progenitor detections are made. Subtle differences in bandwidth, filter profile and effective wavelength of the different $I$-band filters mean that they cannot be directly compared to one another, and so colour transformations must be made (see Appendix A). 

We note that these values may contain some systematic errors. For example, the poorer sampling of the later spectral types, caused presumably by the shorter amount of time that stars spend at these spectral types, may result in the absolute value of the BCs being systematically underestimated. This would ultimately cause the progenitor masses to be underestimated in the case of a detection in any band shortward of $I$. Since this is the case for the vast majority of the SNe in our sample, the ultimate consequence would be that the upper mass cutoff \mhi\ would be underestimated. 




\begin{table*}
\caption{The SNe in our sample, along with their adopted distance moduli, pre-explosion brightnesses and bandpass, extinctions, adopted BCs, inferred luminosities and initial masses. See text for details.}
\begin{center}
\setlength{\extrarowheight}{6pt}
\begin{tabular}{lccccccc}
\hline \hline
id & $\mu$ & $m_\lambda$ & $\lambda$ & $A_\lambda$ & {\it BC}$_\lambda$ & $\log(L/L_\odot)$ & $M_{\rm init}/M_\odot$ \\
\hline
SN1999an & 31.34$\pm$0.08 &       $>$24.70 &      F606W &  0.28$\pm$0.13 & -1.83$\pm$0.38 & $<$5.40$^{+0.11}$ & $<$21.9$^{+3.2}$ \\ 
SN1999br & 30.75$\pm$0.18 &       $>$24.90 &      F606W &  0.04$\pm$0.04 & -1.83$\pm$0.38 & $<$4.98$^{+0.12}$ & $<$14.1$^{+1.8}$ \\ 
SN1999em & 30.34$\pm$0.09 &       $>$23.00 &      $I_C$ &  0.16$\pm$0.08 & -0.32$\pm$0.15 & $<$5.02$^{+0.06}$ & $<$14.7$^{+0.9}$ \\ 
SN1999gi & 30.00$\pm$0.08 &       $>$24.90 &      F606W &  0.45$\pm$0.11 & -1.83$\pm$0.38 & $<$4.85$^{+0.10}$ & $<$11.8$^{+2.0}$ \\ 
SN2001du & 31.31$\pm$0.07 &       $>$24.25 &      F814W &  0.26$\pm$0.14 &  0.00$\pm$0.15 & $<$4.82$^{+0.07}$ & $<$11.4$^{+1.2}$ \\ 
SN2002hh & 28.85$\pm$0.07 &       $>$22.80 & $i^\prime$ &  2.60$\pm$0.10 & -0.49$\pm$0.15 & $<$5.55$^{+0.06}$ & $<$26.3$^{+1.6}$ \\ 
SN2003gd & 29.84$\pm$0.19 & 24.00$\pm$0.04 &      F814W &  0.23$\pm$0.07 &  0.10$\pm$0.15 & 4.28$\pm$0.09 &  6.4$^{+0.6}_{-0.4}$ \\ 
 SN2004A & 31.54$\pm$0.17 & 24.36$\pm$0.12 &      F814W &  0.34$\pm$0.10 &  0.00$\pm$0.15 & 4.90$\pm$0.10 & 12.7$^{+1.6}_{-1.5}$ \\ 
SN2004dg & 31.51$\pm$0.13 &       $>$25.00 &      F814W &  0.40$\pm$0.05 &  0.00$\pm$0.15 & $<$4.66$^{+0.07}$ & $<$ 9.5$^{+0.7}$ \\ 
SN2004et & 28.85$\pm$0.07 & 22.06$\pm$0.12 &      $I_J$ &  0.64$\pm$0.05 &  0.25$\pm$0.15 & 4.77$\pm$0.07 & 10.7$^{+0.9}_{-0.8}$ \\ 
SN2005cs & 29.62$\pm$0.12 & 23.62$\pm$0.07 &      F814W &  0.27$\pm$0.03 &  0.05$\pm$0.15 & 4.38$\pm$0.07 &  7.1$^{+0.5}_{-0.5}$ \\ 
SN2006bc & 30.84$\pm$0.18 &       $>$24.45 &      F814W &  0.34$\pm$0.00 &  0.00$\pm$0.15 & $<$4.59$^{+0.08}$ & $<$ 8.8$^{+0.8}$ \\ 
SN2006my & 31.74$\pm$0.12 & 24.86$\pm$0.13 &      F814W &  0.81$\pm$0.42 &  0.00$\pm$0.15 & 4.97$\pm$0.18 & 13.9$^{+2.9}_{-3.0}$ \\ 
SN2006ov & 30.50$\pm$0.19 &       $>$24.20 &      F814W &  0.13$\pm$0.04 &  0.00$\pm$0.15 & $<$4.47$^{+0.08}$ & $<$ 7.8$^{+0.7}$ \\ 
SN2007aa & 31.56$\pm$0.13 &       $>$24.44 &      F814W &  0.05$\pm$0.00 &  0.00$\pm$0.15 & $<$4.76$^{+0.06}$ & $<$10.6$^{+0.8}$ \\ 
SN2008bk & 27.96$\pm$0.13 & 18.39$\pm$0.03 &          K &  0.03$\pm$0.01 &  3.00$\pm$0.18 & 4.53$\pm$0.07 &  8.3$^{+0.6}_{-0.6}$ \\ 
SN2008cn & 32.61$\pm$0.10 & 25.13$\pm$0.09 &      F814W &  0.54$\pm$0.06 &  0.00$\pm$0.15 & 5.10$\pm$0.07 & 15.9$^{+1.2}_{-1.1}$ \\ 
SN2009hd & 29.86$\pm$0.08 & 23.54$\pm$0.14 &      F814W &  2.04$\pm$0.08 &  0.00$\pm$0.15 & 5.24$\pm$0.08 & 18.3$^{+1.6}_{-1.5}$ \\ 
SN2009kr & 32.09$\pm$0.50 & 24.71$\pm$0.23 &      F555W &  0.35$\pm$0.03 & -0.36$\pm$0.20 & 5.13$\pm$0.23 & 16.4$^{+4.6}_{-3.6}$ \\ 
SN2009md & 31.64$\pm$0.21 &           -    &      multi &           -    &           -    & 4.50$\pm$0.20 &  8.0$^{+1.9}_{-1.5}$ \\ 
 SN2012A & 29.96$\pm$0.15 & 20.29$\pm$0.13 &          K &  0.01$\pm$0.00 &  3.00$\pm$0.18 & 4.57$\pm$0.09 &  8.6$^{+0.9}_{-0.8}$ \\ 
SN2012aw & 29.96$\pm$0.02 & 19.56$\pm$0.29 &          K &  0.16$\pm$0.02 &  3.00$\pm$0.18 & 4.92$\pm$0.12 & 13.0$^{+1.9}_{-2.0}$ \\ 
SN2012ec & 31.19$\pm$0.10 & 23.39$\pm$0.08 &      F814W &  0.36$\pm$0.10 &  0.00$\pm$0.15 & 5.16$\pm$0.07 & 16.8$^{+1.4}_{-1.3}$ \\ 
SN2013ej & 29.80$\pm$0.11 & 22.65$\pm$0.05 &      F814W &  0.24$\pm$0.03 &  0.40$\pm$0.20 & 4.69$\pm$0.07 &  9.8$^{+0.8}_{-0.7}$ \\ 

\hline
\end{tabular}
\end{center}
\label{tab:results}
\end{table*}%

\section{Reappraisal of progenitor masses} \label{sec:prog}
In the previous sections we have shown that RSGs systematically evolve to later spectral types as they approach SN, with a substantial impact on their BCs. With this in mind, we now take a fresh look at the inferred progenitor masses for all SNe with pre-explosion detections and/or upper limits.

\subsection{Sample}
For the sample of objects, we begin with those events listed in S09 and S15. The mass-spectrum of II-P/II-L progenitors was first defined in S09, but some of the terminal luminosities (and therefore initial masses) of individual events were revised in S15 by adding in an ad-hoc level of circumstellar extinction. S15 also added several events that had occurred in the intervening years. Unless stated otherwise in Sect.\ \ref{sec:individ}, we adopt the same distance moduli and pre-explosion photometry as in S15 (and references therein). 

For any event with a single-band pre-explosion detection, we have revised the estimate of the terminal luminosity assuming a spectral type of M4-M7, as discussed in Sect.\ \ref{sec:bc} and listed in Table \ref{tab:bcs}. For certain events, information from multiple bands allows us to place constraints the terminal colour of the progenitor. We have not revised the pre-explosion luminosity for SN2009md, where multi-band detections have already tightly constrained the colour of the progenitor.  Other events have single-band detections plus upper-limits in other bands which can constrain the terminal spectral type (e.g. SN2013ej). Below we describe the state of the observations for each SN in the sample where the observed quantities have been updated since S09/S15, and our assumptions which lead to progenitor mass estimates for these objects. Where possible, the foreground reddening has been updated to that estimated from the colours of the progenitor's neighbouring stars \citep{Maund14,Maund17}. Objects for which we use the same distance modulus $\mu$, extinction, and progenitor photometry as S09/S15 are not discussed. The observed quantities, and inferred pre-explosion luminosities and initial masses are listed in Table \ref{tab:results}.

\subsubsection{Notes on individual objects} \label{sec:individ}

\paragraph*{SN2003gd -- }  Late-time/pre-explosion difference imaging was used to constrain the progenitor's pre-SN brightness to be $m_{\rm F814W} = 24.90\pm0.04$ \citep{Maund14}. Detections in two bluer bands constrain the pre-SN spectral type to be M3 or earlier \citep{Smartt04,Maund14}. We have assumed a spectral-type of M3, with $BC_{\rm F814W} = 0.10 \pm 0.15$, using our results from \fig{fig:spt-bc} and \fig{fig:colours}. \citet{Maund14} estimated the foreground extinction to this object using the colours of the neighbouring stars to be E(B-V)=0.14$\pm$0.04, hence $A_V=0.45\pm0.19$ using the extinction law of Sect.\ \ref{sec:extinct}.

\paragraph*{SN2004A -- } This progenitor was detected in a pre-explosion WFPC2 F814W image, but not detected in F606W. \citet{Maund14} used late-time/pre-explosion imaging to refine the pre-explosion brightness to be $m_{\rm F814W}$ = 24.36$\pm$0.12. The estimated foreground extinction is $A_V=0.64\pm0.19$ \citep{Maund17}, which corresponds to $A_{\rm F814W} = 0.34\pm0.10$ when adopting the extinction law of Sect.\ \ref{sec:extinct}.

\paragraph*{SN2004et -- } There has been some disagreement in the literature about the progenitor to this SN (see S09 for discussion). In this analysis we follow the results of \citet{Crockett11}, who argue that the progenitor star is that detected at $i^\prime$, with useful upper limits in $BVR$ that confine the spectral type to M4 or later. The $i^\prime$ flux was converted to the Johnson-$I$ by S09. We adopt the foreground extinction of $A_V = 1.2 \pm 0.1$ ($A_{I} = 0.64\pm0.05$), and the BC$_I$ for a late-type RSG.

\paragraph*{SN2005cs -- } \citet{Maund05} present a multi-wavelength study of this progenitor in the {\it BVRIJHK} bands. Unfortunately, the only detection is in the $I$-band, with the upper limits in the other bands constraining the spectral type to within K5-M4. \citet{Maund14} use difference imaging to tune-up the measured progenitor brightness, finding $m_{\rm F814W}$=23.62$\pm$0.07. The neighbouring stars imply a foreground extinction of $A_V=0.54\pm0.06$ ($A_{\rm F814W} = 0.27\pm0.03$). We adopt the BC of an M4 RSG.

\paragraph*{SN2006my -- } Though there had previously been disagreement in the literature as to whether the pre-explosion source was detected, \citet{Maund14} used difference-imaging to constrain the pre-explosion brightness to be $m_{\rm F814W}$=24.86$\pm$0.13, and inferred a reddening from the neighbouring stars \ebmv=0.49$\pm$0.25 ($A_{\rm F814W} = 0.81\pm0.42$).

\paragraph*{SN2008bk -- } The pre-explosion photometry for this progenitor includes detections in several bands from the optical to the near-IR. Further, late-time imaging of the SN site by \citet{Maund14bk} confirms the progenitor is gone, and allows for precise differential photometry in each of the detection bands. By modelling the SED, these authors were able to place estimates on the pre-explosion luminosity (and hence progenitor initial mass), though without an independent estimate of the foreground extinction there was a degeneracy between initial mass and \ebmv. \citet{Maund17} modelled the surrounding population and measured this foreground extinction to be much lower than had previously been accounted for -- $A_V=0.28\pm0.10$, whereas the estimate in \citet{Maund14bk} was almost $\times$10 higher. This new extinction estimate therefore implies a much redder progenitor than previously thought. 

To estimate the pre-explosion luminosity, we have taken the $K_S$-band difference-image brightness of $m_{Ks} = 18.39\pm0.03$, and applied the extinction from \citet{Maund17} scaled via our extinction law to the $K$-band, $A_K = 0.03\pm0.01$, and used the $K$-band BC. We note that if a $K$-band pre-explosion detection exists, the relative insensitivity of the flux at this wavelength to \BCK\  means that accurate luminosities can be estimated without the need for modelling the SED \citep[see ][ for a demonstration of this using SN2008bk as an example]{rsgteff}. With the updated extinction of \citet{Maund17}, we now find a somewhat lower progenitor mass for this object (8\msun\ vs 11-12\msun).

\paragraph*{SN2008cn -- }  This SN was not included in the S15 sample. Following \citet{Elias-Rosa09}, we adopt the cepheid distance modulus of $\mu$ = 32.61 $\pm$ 0.10mag \citep{Newman99}, and a foreground extinction of \ebmv=0.35$\pm$0.04 which was determined from the equivalent width of the Na~{\sc i}~D lines in the SN spectrum. \citet{Maund15} measure $m_{\rm F814W}$ = 25.13 $\pm$ 0.09 from difference imaging, though they remark that the pre-explosion source may be a blend of two or more stars, in which case this would be an upper limit. For our analysis we have assumed that the pre-explosion flux comes entirely from the progenitor.

\paragraph*{SN2009kr -- } The detection of this SN's progenitor in the F555W and F814W bands led to the conclusion that it was a yellow supergiant, with a $V-I$ colour consistent with mid-G type \citep{Fraser10}. Assuming foreground extinction from the Milky Way only ($A_V=0.242$), these authors determined a pre-explosion luminosity of \logl=$5.1\pm0.24$ (and hence $M=15^{+5}_{-4}$\msun). The error on this value is dominated by that on the distance to the host galaxy. 

\citet{Maund15} re-analysed the progenitor by comparing late-time and pre-explosion imaging. They claimed that the pre-explosion source is in fact a cluster, age analysis of which yields a progenitor mass estimate between 12-25\msun. 

In this work, we have adopted the luminosity estimate of \citet{Fraser10}. However, we note that the mass constraints of \citet{Maund15} are very similar. This means that the impact on the inferred upper limit to the progenitor mass distribution is only very minor, regardless of whether the Maund or Fraser progenitor information is used (see Sect.\ \ref{sec:mdist}).

\paragraph*{SN2009hd -- } This SN was originally studied in \citet{Elias-Rosa11}. A progenitor candidate was detected at the SN site in the F814W WFPC2 image, with a brightness of $m_{\rm F814W} = 23.54 \pm 0.14$, though the authors note that this is close to their empirically-derived detection limit. No progenitor candidate is detected at F555W, and the upper limit places no useful constraint on the progenitor's colour. The authors estimate the progenitor luminosity \lfin\ by taking the F555W upper limit, correcting for a large foreground extinction of $A_V=3.80\pm0.14$, obtained from analysis of the SN spectrum, and assume a BC appropriate for an early K supergiant of -0.29mag. 

In our analysis, we take the F814W detection, correct for the same extinction scaled to the F814W passband ($A_{\rm F814W} = 2.04\pm0.08$), and assume a late-type RSG progenitor. This results in a substantial increase in the star's \lfin, going from $\log(L/L_\odot) = 5.04$ to $5.30\pm0.08$.

\paragraph*{SN2009md -- } Progenitor detections in the HST filters F606W and F814W are presented in \citet{Fraser11}, and the pre-explosion colour implies a spectral type of M4, consistent with our earlier conclusions that RSGs explode with late spectral types. However, \citet{Maund15} find that the brightness of the source at the SN location in late-time images is roughly the same as that in the pre-explosion images., calling into question the conclusions of \citet{Fraser11}.  

For the purposes of this work we have adopted the \citet{Fraser11} pre-explosion brightness at F814W, with the BC appropriate for a late-type RSG, with the foreground extinction from the stars local to the SN site \citep{Maund17}. Though questions still persist over the nature of this progenitor, we note that even with the measurements of Fraser et al.\ the inferred progenitor mass is low ($\sim$8\msun), so the impact of this object on the inferred upper mass cutoff is minimal.

\paragraph*{SN2012A -- } This SN has a single-band detection of the progenitor in a Gemini/GNIRS $K'$-band image \citep{Tomasella13}. We adopt the same distance and foreground extinction as these authors, the latter derived from the Na~{\sc i}~D lines in the SN spectrum. With no colour information on the progenitor, we again assume a late-type RSG.

\paragraph*{SN2012aw -- } The progenitor mass of this object has been subjected to scrutiny by several authors in the literature. The original estimates by \citet{Fraser12} and \citet{vanDyk12} suggested mass-ranges of 14-26\msun\ and 15-20\msun\ respectively. Both studies included a substantial amount of circumstellar extinction in their analysis. However, \citet{Kochanek12} argued that the reduction in the optical flux from this extinction was overestimated, since circumstellar dust can scatter photons {\it in} to the line-of-sight as well as out of it, and revised the progenitor mass down to $<15$\msun. The mass of the progenitor was revisited by \citet{Fraser16}, using late-time imaging to perform accurate differential photometry between pre-  and post-explosion. Detections were found in {\it JHK}, which led to an inferred pre-explosion luminosity of \logl=4.8-5.0, consistent with Kochanek et al. 

By using {\it empirical} bolometric corrections, we should be immune to the specifics of how efficient circumstellar dust is at extinguishing optical light, since we are comparing {\it observed} photometry with {\it observed} luminosities. Also, by using the $K$-band detection, the luminosity estimate is largely insensitive to spectral type (see \fig{fig:spt-bc}). We adopt the extinction derived from the colours of the stars local to the SN site of $A_V=1.37\pm0.2$ \citep{Maund17}, and find a progenitor mass consistent with that in \citet{Kochanek12} and \citet{Fraser16} (13$\pm$2\msun, see Table \ref{tab:results}).

\paragraph*{SN2012ec -- } The progenitor had a detection in the F814W filter and a useful upper limit at F555W, which limited the spectral type to be later than $\sim$K0 \citep{Maund13}. We again assume a late-type RSG progenitor, and the extinction of \citet{Maund17} ($A_V=0.72\pm0.2$).

\paragraph*{SN2013ej -- } This object was detected in F555W and F814W, with its terminal colour indicating a spectral type of M2 \citep{Fraser14}. We adopt a BC appropriate for this spectral type based on our results in Sect.\ \ref{sec:bc} of \BCI=0.4$\pm$0.2, and the extinction measured from the neighbouring stellar population of $A_V = 0.45 \pm 0.06$ \citep{Maund17}.

\begin{figure}
\begin{center}
\includegraphics[width=8.5cm,bb=50 10 546 500,clip]{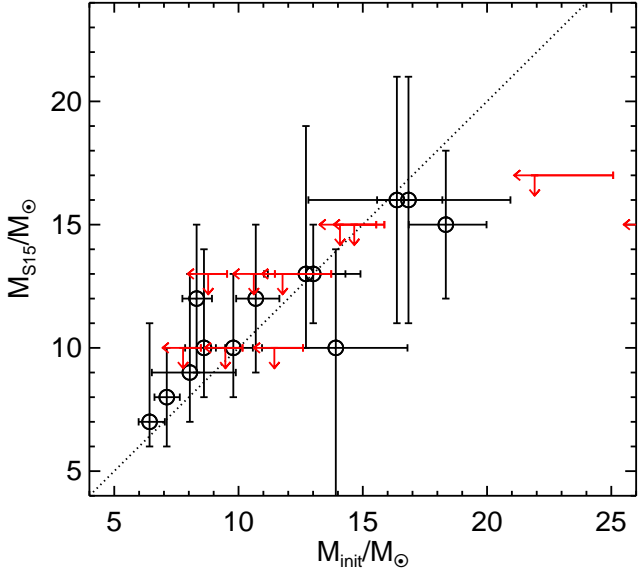}
\caption{Comparison of the progenitor masses derived in this study with those from S15.}
\label{fig:mass-comp}
\end{center}
\end{figure}

\subsection{Comparison with previous mass estimates}
Our updates to the progenitor masses are listed in the final column of Table \ref{tab:results}. These are compared with the masses estimated in S15 \fig{fig:mass-comp}. On average, the progenitor masses have increased with respect to S15. For many objects, the change in mass is small, as the updated BCs in this work are somewhat cancelled out by the ad hoc inclusion of $A_V=0.5$mag of circumstellar extinction in S15. The objects with pre-explosion detections in $K$ (SN2012A, SN2012aw) have all moved to slightly lower \lfin\ due to the larger positive BC at this wavelength, while SN2008bk has changed dramatically due to the downward revision in its foreground extinction. The objects with notable increases in progenitor mass are SN2006my, SN2012ec and SN2009hd. In the latter two SNe, these shifts are due not just to the change in BC, but also to the foreground extinction being revised upwards from analysis of the local stellar population. In the case of SN2009hd, the shift to higher progenitor mass is due to the large change in BC compared to that assumed by \citet{Elias-Rosa11}.

\begin{figure}
\centering
\includegraphics[width=8.5cm,bb=20 20 553 463,clip]{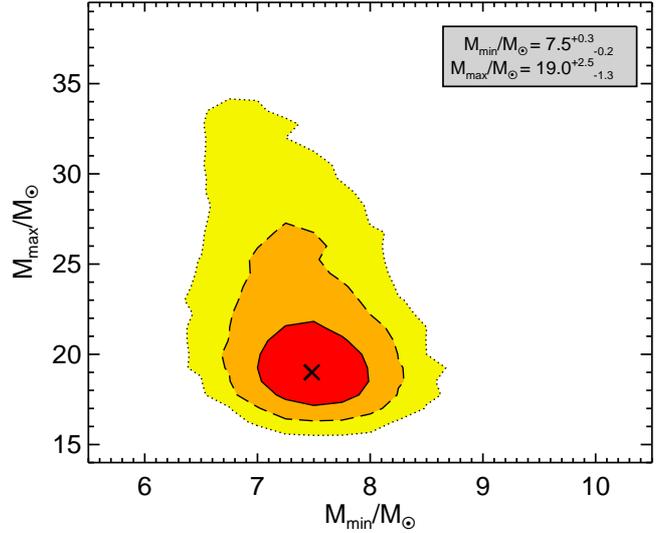}
\caption{Results from the MC analysis of the luminosity distributions of the progenitor stars. The left panel uses the luminosities from S09, and the centre panel from S15. The right panel uses bolometric luminosities derived from our empirically-motivated BCs. }
\label{fig:mlo-mhi}
\end{figure}

\begin{figure}
\begin{center}
\includegraphics[width=8.5cm,bb=35 10 663 503,clip]{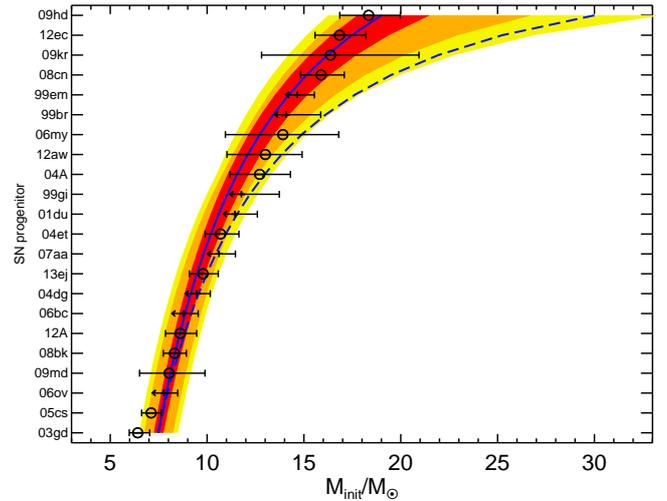}
\caption{The cumulative mass function of the SN progenitors, using masses inferred from our updated BCs and the extinction estimates of \citet{Maund17}. The solid blue line is the best-fit, the red/orange/yellow shaded regions denote all solutions within 1/2/3$\sigma$ respectively. The dashed line is the mass function assuming a Salpeter IMF and an upper mass limit of 30\msun.}
\label{fig:mass-spec}
\end{center}
\end{figure}

\subsection{Estimating the upper and lower mass limits to the progenitor distribution} \label{sec:mdist}
The conversion of pre-explosion photometry to bolometric luminosity, and ultimately to initial mass, involves the combination of errors which can be non-Gaussian, non-linear, and asymmetric. It is non-trivial to propagate these errors in an analytic way. We therefore opt to fit the mass spectrum of the progenitors using a Monte-Carlo (MC) method, which we now describe. 

For each MC trial, we begin by sampling the pre-explosion photometry of each progenitor from a Gaussian distribution centred on the measured magnitude $m_\lambda$ with a 1$\sigma$ width defined by the error on $m_\lambda$ quoted in the relevant paper. This is converted to an absolute magnitude by sampling the distance modulus from a Gaussian distribution. We correct for foreground extinction by again sampling from a Gaussian distribution, but with a cutoff at zero since negative extinctions are unphysical. The extinction at the detected waveband has been determined using reddening measurements in the literature and our extinction law described in Sect.\ \ref{sec:extinct}. The absolute magnitude is then converted to a bolometric magnitude by sampling $BC_{\lambda}$ from a uniform (flat) distribution between the limits in column 6 in Table \ref{tab:results}. We opt for a flat distribution that defines a range of allowed BC values, rather than Gaussian distribution with a mean and a variance, since the latter better reflects our uncertainty on the precise value of BC to use. The bolometric magitude is converted to a bolometric luminosity assuming an absolute magnitude of the Sun of 4.74 mag. 

The final step is to convert the pre-explosion bolometric luminosity to an initial mass. To do this, we use the exact same models as those employed in S09 and S15, specifically the STARS models \citep{Eldridge08}, which describe the relationship between initial mass and the luminosity of the star (hereafter the mass-luminosity relation, or MLR) at the beginning of Ne burning, after which the star only has a matter of weeks to live\footnote{The sensitivity of our results to the MLR are explored in Sect.\ \ref{sec:mlr}. Also, we note that the MLRs we explore are valid only for single stars, and do not apply to stars with a history of interaction with a companion.}. At this evolutionary phase the stars can have systematically higher luminosities than at the end of He or C burning, where most published evolutionary tracks end. Since the published MLR stops at a mass of 25\msun, for luminosities higher than this limit we extrapolate the mass from a linear fit between $\log(M)$ and $\log(L)$, specifically $\log(M/M_{\odot} = -1.2417 + 0.47868\log(L/L_{\odot})$. This fit describes the MLR to within 0.5\msun\ between the limits of 8\msun\ to 25\msun, but clearly there is a potential source of systematic error at higher extrapolated masses.  

For progenitors with only upper limits to their pre-explosion brightness, we take the quoted 5$\sigma$ detection limit on the photometry, and for each trial convert this into upper limits on the luminosity and initial mass following the same procedure as with the detections. This means that in our analysis the upper limits have an effective probability distribution function which is unity everywhere below the upper limit, with a Gaussian-like tail above it. 

In each MC trial, the above step is repeated for each progenitor. The masses and the upper limits are then sorted in increasing order, and fit with the following function which determines the initial mass $M_n$ of the $n$th progenitor,

\begin{eqnarray}
\mathcal{N} & = & \frac{1}{M_{\rm hi}^\Gamma - M_{\rm lo}^\Gamma} \\
\mathcal{E} & = & \frac{1}{\Gamma} \log( n/\mathcal{N} + M_{\rm lo}^\Gamma) \\
M_{n} & = & 10^\mathcal{E}
\end{eqnarray}

\noindent using the IDL $\chi^2$ minimisation routine {\tt curvefit}. The parameter $\Gamma$ is the slope of the initial mass function, and is fixed at -1.35\footnote{In principle we could also allow $\Gamma$ to be a variable, however the fitted value of $\Gamma$ has no effect on the parameter we are most interested in, \mhi.}. The parameters \mlo\ and \mhi\ are the upper and lower mass limits for the progenitor distribution respectively, and are allowed to be free. Each MC trial produces values of \mlo\ and \mhi. The progenitors with upper limits are not explicitly fitted -- we fit only to those values of $n$ that have a detection. The information carried by the upper limits is therefore the following: they tell us that in a given trial, for the $n$th SN with progenitor mass $M_n$, there are ($n$-1) SNe with lower progenitor masses. Since there is no diagnostic power in any upper limit greater than the highest mass detection, in any trial where this is the case these high upper limits are excluded from the fit. We ran $10^5$ trials, and binned the values of \mlo\ and \mhi\ in a two-dimensional histogram. 

The results of this analysis are shown in \fig{fig:mlo-mhi}. The best-fit values of \mlo\ and \mhi\ are found by compiling histograms of their best-fit values in each trial, and linearly interpolating the point at which the derivatives of these histograms goes to zero. Error contours analogous to 1, 2 and 3 $\sigma$ uncertainties were found by determining the regions which contained 68\%, 95\% and 99.7\% of the trial results. We find that the lower mass limit is tightly constrained at \mlo=$7.5^{+0.3}_{-0.2}$\msun. The best-fitting upper mass limit is \mhi=$19.0^{+2.5}_{-1.3}$\msun. The 95\% confidence limit is \mhi$<27$\msun. This is somewhat higher than the S15 value of \mhi=16.5$\pm$2.5\msun, while our 95\% confidence limit is substantially higher than that quoted by  S15 (22\msun). We note that, if we use the S15 luminosities, we also recover the same \mhi\ as S15, but with a slightly higher 95\% confidence limit of 24\msun. Finally, we add that we recover the same values of \mhi\ and \mlo\ to within the errors whether or not the upper limits are included in the fit. 

To provide an illustration of our result and our error margins, in \fig{fig:mass-spec} we plot a cumulative IMF using the SNe in our sample. The error bars on the progenitor mass of each SN represent the 68\% confidence limits from the MC trials. The progenitors are ordered in increasing mass, and the black solid line shows the IMF with our best-fitting \mlo\ and \mhi. The dashed line shows the IMF for the same \mlo\ but an upper mass cutoff of 30\msun. What the plot illustrates is that, though the best fit \mlo\ is still lower than the 25-30\msun\ expected from theory, the significance of this disagreement is, at best, 2.5$\sigma$.

\subsubsection{The sensitivity of \mhi\ on the few high-mass progenitors}
What \fig{fig:mass-spec} also helps to demonstrate is the sensitivity \mhi\ on the small number of progenitors with large masses. The exact value of \mhi\ is strongly influenced by the most massive object in the sample. Meanwhile, the error on \mhi\ is largely governed by the errors on the masses of the $\sim$5 most massive objects. For example, if late-time imaging of SN2009hd were to reveal that the putative progenitor was still there, and that this object was either removed from the sample or had its mass revised substantially downwards, the best-fit value of \mhi\ would move closer 18\msun\ -- the mass of the next most massive object in the sample (SN2012ec). However, the error on \mhi\ would still be large, as the size of this error is driven mostly by the relatively large errors on SN2009kr and SN2006my. If we follow \citet{Maund15} and say that, rather than having a YSG progenitor, the progenitor of SN2009kr was a star in an unresolved cluster, and use the age estimates of this cluster to constrain the progenitor mass, the similarly of the permitted mass range to that inferred for the YSG progenitor by \citet{Fraser10} means that the effects on \mhi\ and its error are minimal. If SN2009kr was removed from the sample altogether, the error on \mhi\ would remain roughly the same due to the similarly large error on the progenitor mass of SN2006my. The only way to substantially reduce the error on \mhi\ is to remove SN2009kr and SN2006my from the sample. If we do remove these two SNe, \mhi\ stays the same to within 0.5\msun, but the 95\% confidence limit reduces to 24\msun. However, the tension with theory remains weak ($\sim 3 \sigma$). 

To summarise, the only way to reduce the best-fit value of \mhi\ is to remove or revise downwards the progenitor mass of SN2009hd. The only way to reduce the error on \mhi\ is to disregard the SNe SN2009kr and SN2006my. In the future, when the number of SNe with progenitor detections is much larger, excluding the datapoints with large errors may be an effective way to reduce the errors on the limits to the mass distribution. However, with the current sample size of progenitor detections standing at just 14, discarding data-points based on their error bars may introduce significant systematic errors. For example, the most massive progenitors should have shorter lifetimes, and hence be more likely to explode in regions of high stellar density. As such, the higher mass progenitors may be more likely to have larger error bars on their pre-explosion brightness, and ultimately their mass. If this were the case, discarding points with large errors would skew \mhi\ to lower masses.

\begin{figure}
\begin{center}
\includegraphics[width=8.5cm]{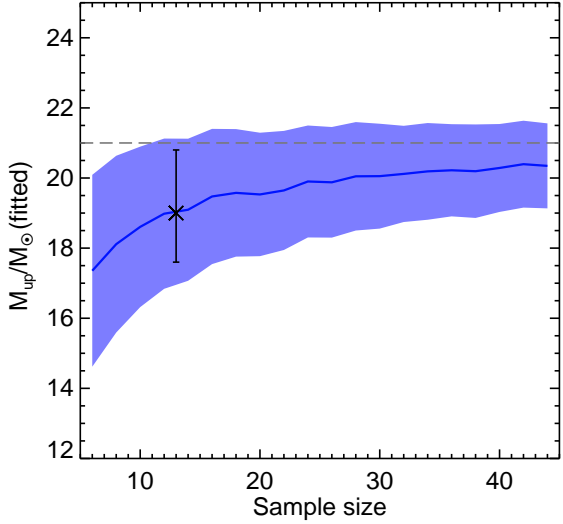}
\caption{The average measured upper mass limit as a function of sample size (blue line) for an intrinsic upper mass limit of 21\msun. The shaded region shows the 1$\sigma$ dispersion on this average. The data point is the result from this study. }
\label{fig:finite}
\end{center}
\end{figure}

\subsubsection{The effect of a finite sample size}
As already mentioned in the previous Section, the upper mass limit inferred is strongly dependent on the most massive object in the sample. For a very large sample, the mass of the most massive object will be very close to the true upper mass limit, and so the inferred upper mass limit will be close to the true value. However, as we are sampling from a distribution with a steep negative power law, as one decreases the size of the sample the likelihood that the most massive object in the sample will be close to the true upper mass limit decreases. This means that, in a finite sample, one will always systematically underestimate the upper mass cutoff, and this systematic error will be larger for smaller sample sizes. 

To understand the nature of this systematic error we have performed a simple numerical experiment. We randomly generated a sample of masses of size $\mathcal{N}$ drawn from an IMF-like distribution with a fixed upper and lower mass limit. We then performed the same analysis as described above to determine the {\it observed} mass limits. For each size of sample, we repeated the experiment $10^4$ times to determine the most likely value for that sample size as well as the dispersion. The results are shown in \fig{fig:finite}. Broadly, the plot shows what one would expect: that the inferred upper mass limit is always below the true upper mass limit, and that this effect is larger for smaller $\mathcal{N}$. The datapoint on the plot shows our measurement from the previous section; a sample size of 14 detections and an inferred \mhi=19\msun. The result of this experiment suggests then that our small sample induces a systematic error on \mhi\ of 2\msun, and that the most likely true upper mass limit is 21\msun.

\subsubsection{The impact of uncertainties in the mass-luminosity relation} \label{sec:mlr}
The adopted MLR is a vital component in the conversion of the pre-explosion luminosity into an initial stellar mass. In the work presented here, up to now we have used the same MLR as used by S09/S15. By keeping as many components as possible the same between our work and S09/S15, we have been able to study the impact on \mhi\ of more realistic BCs and foreground extinctions. However, the choice of MLR also represents a major source of systematic error, as we will now show. 

In stellar evolution models, internal mixing processes -- whether due to rotation, convection, semi-convection, or convective overshooting -- alter the chemical composition of the core, and ultimately impact the star's luminosity in the post-MS phases. As shown by \citet{Jerkstrand14} in their Fig.\ 5, the difference between the pre-SN MLR of three competing evolutionary codes (specifically the STARS models used here, the Geneva models of \citet{Ekstrom12}, and the KEPLER models of \citet{Woosley-Heger07}) can be as large as 20\% in terms of mass. That is, a luminosity of $\log(L/L_\odot) = 5.35$ would imply a 20\msun\ progenitor according to STARS, but would be a 24\msun\ according to KEPLER. 

To quantitatively investigate the impact of the choice of MLR on \mhi, we re-ran our analysis using the MLR of KEPLER \citep[shown in Fig.\ 5 of][]{Jerkstrand14}. While the lower mass cutoff moves only slightly to \mlo=8.7$^{+0.6}_{-0.4}$\msun, the upper mass cutoff moves up to \mhi=$24.2^{+3.3}_{-1.9}$\msun, with a 95\% confidence upper limit of 32\msun. What this clearly shows is that, once uncertainties in the MLR are taken into account, the evidence for a disagreement between observation and theory over the value of \mhi\ has very low statistical significance.

\begin{figure}
\begin{center}
\includegraphics[width=8.5cm]{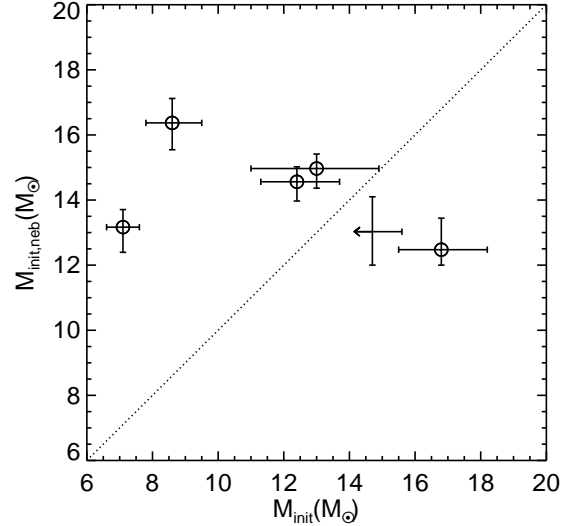}
\caption{A comparison of the masses from our study with those derived from estimates of the O abundance in the ejecta from late-time `nebular' spectroscopy.}
\label{fig:jerk}
\end{center}
\end{figure}

\subsubsection{Comparison with progenitor mass estimates from late-time nebular spectra}
In this Section we compare our progenitor masses with those inferred from analysis of [O{\sc i}] lines in late-time spectra. These lines have been used to estimate the mass of O in the ejecta which itself is thought to be sensitive to the initial mass of the progenitor. By making model predictions about the flux in the optical [O{\sc i}] lines as a function of time since explosion, a progenitor mass may then be estimated \citep{Jerkstrand14,Jerkstrand15}. 

Such mass estimates in the literature tend to merely place upper and lower limits based on which models can be ruled out. To get more accurate mass estimates, we have taken the [O{\sc i}] line-flux measurements for a sample of SNe, and compared them quantitatively to the model predictions. The observations and models are taken from \citet{Valenti16}, and include observations of five SNe common with the sample in our current work. We first took the models, which predict the [O{\sc i}] as a function of time for a range of initial progenitor masses, and interpolated them onto a finer grid in mass. For each SN in the sample, we computed the $\chi^2$ at each interpolated progenitor mass $m$,  

\begin{equation}
\displaystyle
\chi^2 = \sum_t \left(\frac{f^{\rm obs}_{t} - f^{\rm mod}_{t}}{\sigma^{\rm obs}_{t}}\right)
\end{equation}

\noindent where $f^{\rm obs}_{t}$ is the observed [O{\sc i}] line-fluxes as a function of time $t$, $f^{\rm mod}_{t}$ is the model line-flux at the same time at a progenitor mass $m$, and  $\sigma^{\rm obs}_{t}$ is the observational error on $f^{\rm obs}_{t}$. The best fitting progenitor mass $M_{\rm init,neb}$ was found by first taking the derivative of the $\chi^2$ with respect to $m$, and interpolating to where it crosses zero. The 1$\sigma$ upper and lower error bars were found from the minimum and maximum masses that had $\chi^2 < \chi^2_{\rm min} + 2.3$. 

The comparison of the two sets of mass estimates are shown in Fig.\ \ref{fig:jerk}. The SNe included are SN2004et, SN2005cs, SN2012A, SN2012aw, SN2012ec, and we have also included SN1999em which has an upper limit measurement. It is immediately obvious that there is a lack of correlation between the two independent mass estimates. What is also clear is that, while the distribution of masses estimated from pre-explosion imaging seems to vaguely follow the IMF, the `nebular' mass estimates $M_{\rm init,neb}$ seem to be strongly peaked at $\sim$14\msun. Indeed, the average is $M_{\rm init,neb} = 14.1\pm1.5$\msun. Assuming that the sample of SNe with $M_{\rm init,neb}$ estimates is unbiased, which would seem to be the case based on the $M_{\rm init}$ estimates, this raises some doubts about the accuracy of the $M_{\rm init,neb}$, since the likelihood of have not yet observed a progenitor with a mass $<12$\msun\ out of 6 objects studied is very ($p < 0.06$). 

At present we do not have an explanation for the initial masses from nebular analyses clustering around 14\msun. We speculate two possible solutions: 

\begin{itemize}
\item the [O{\sc i}] line fluxes from progenitors with initial masses lower than the current lowest mass model (12\msun) converge to similar values, meaning that the lower limit error bars on $M_{\rm init,neb}$ in \fig{fig:jerk} are currently underestimated.
\item the high sensitivity of the [O{\sc i}] line fluxes to temperature \citep[as described in][]{Jerkstrand17}, combined with a systematic error in the nebula temperature. Such an error may be caused by macroscopic mixing in the SN ejecta, transporting efficient coolants (e.g. Ca) into the O layer, lowering the temperature and reducing the [O{\sc i}] line luminosities. 
\end{itemize}

\section{Summary \& Conclusions} \label{sec:conc}
We have studied the trend of spectral type and bolometric correction (BC) as a function of evolution among Red Supergiants (RSGs), using resolved young star clusters. We have used our results to take a fresh look at the inferred progenitor masses to nearby type-II supernovae (SNe). Our main findings are as follows:

\begin{itemize}
\item We have shown that RSGs close to SN have much larger absolute BCs than is typically assumed when estimating stellar initial masses of those RSGs identified as SN progenitors from pre-explosion imaging. 
\item We have taken the BCs appropriate for RSGs close to SN and reappraised the initial masses of the progenitors to all SNe with pre-explosion imaging. We have also incorporated updated estimates of the foreground extinction, where possible. We find that, for many of the progenitors, their initial masses are higher than have been previously reported in the literature. 
\item In reanalysing the mass distribution of the progenitors, we find that it is consistent with a standard initial mass function with an upper-mass cutoff of \mhi=19\msun, with a 95\% upper confidence limit of $<$27\msun. 
\item We have investigated two major sources of systematic error on \mhi, specifically those of finite sample size effects and of uncertainties in the pre-SN mass-luminosity relation. These effects account for systematic shifts in \mhi\ of +2\msun\ and +4\msun\ respectively. Taken together, these effects may shift \mhi\ upwards to 25\msun\ (95\% confidence limit of $<$33\msun). 
\item We therefore conclude that there is no strong statistical evidence for `missing' SN progenitors with initial masses $>$17\msun. as has been previously claimed. 
\end{itemize}

\section*{Acknowledgements}
For useful discussions in the course of this work, we would like to thank Rolf-Peter Kudritzki, Nathan Smith, Justyn Maund, and Paul Crowther.




\bibliographystyle{mnras}
\bibliography{/Users/astbdavi/Google_Drive/drafts/biblio} 



\appendix

\section{Conversion between $I$-band fluxes of different photometric systems}
Each photometric system defines its bandpasses differently. This can be particularly problematic when dealing with RSGs in the $I$-band due to the presence of a strong TiO absorption feature. The various common photometric systems sample this feature differently, resulting in non-negligible colour corrections that must be applied when comparing e.g.\ Johnson with Cousins $I$-band magnitudes. This then has important ramifications for the BCs in each photometric system, and as such is an important consideration when attempting to estimate terminal luminosities for SN progenitors. 

\begin{figure}
\begin{center}
\includegraphics[width=8.5cm]{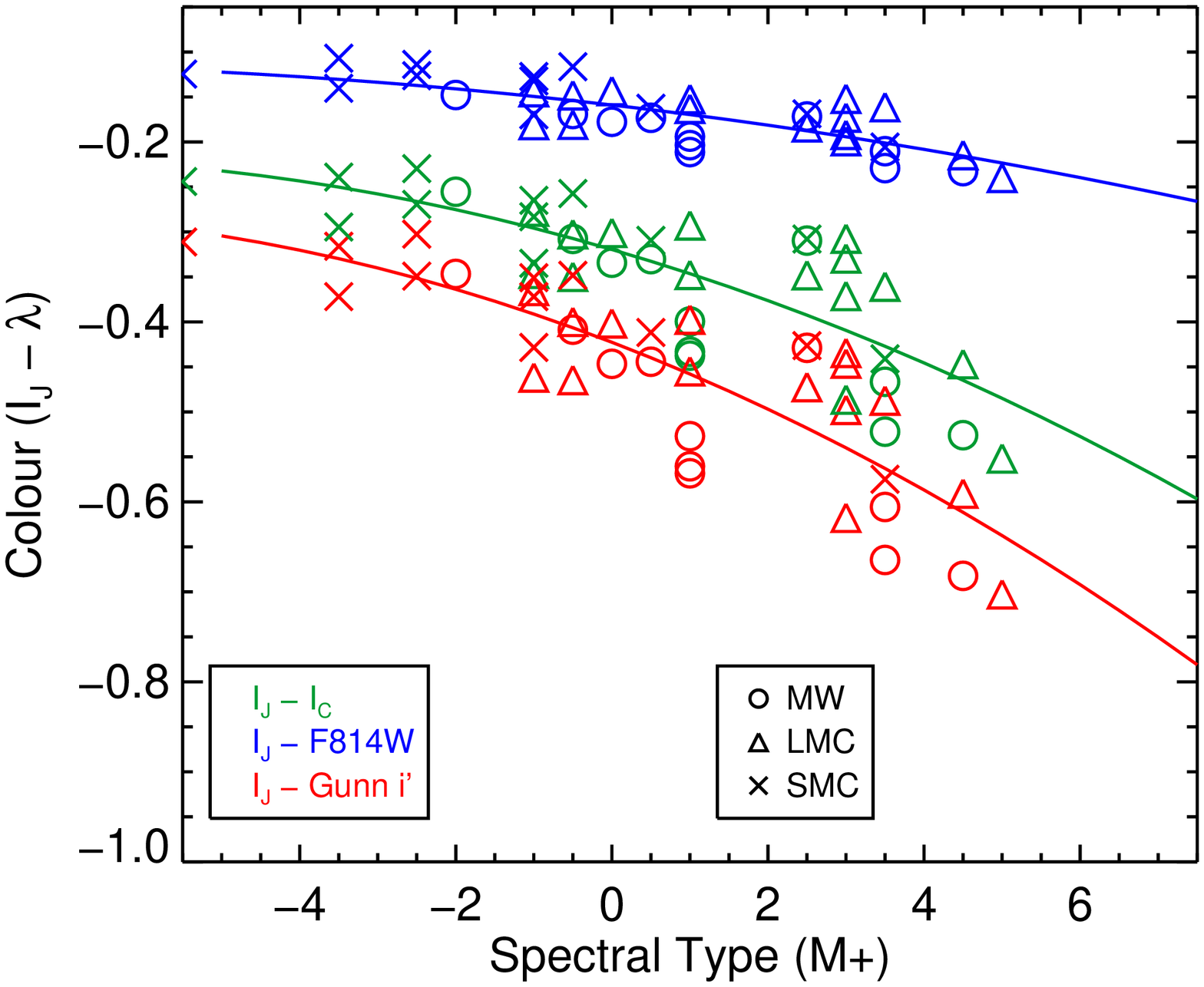}
\caption{The colours for converting between $I$-band fluxes on different photometric systems -- Johnson ($\equiv I_J$), Cousins ($\equiv I_C$), HST-WFPC2 ($\equiv$F814W), and INT/WFC ($\equiv {\it Gunn~i^\prime}$) -- as a function of RSG spectral type. The solid lines are 2nd-order polynomial fits to all data points. The different plotting symbols indicate the host galaxy of each star.}
\label{fig:colours}
\end{center}
\end{figure}

To measure the RSG colour corrections between the different species of $I$-band filter as a function of spectral type, we have assembled a sample of spectrophotometric data for nearby RSGs. These data come from XSHOOTER, and are taken from the XSHOOTER library \citep[XSL, ][, all M supergiants in their database]{Chen14}, and from \citet[][, K/M supergiants]{rsgteff}. The data consists of stars from the Galaxy, the LMC, and the SMC. We have not corrected for any foreground reddening, since we expect the visual extinctions to be small ($\la$1 mag) and so any effect on the relative fluxes in the various incarnations of $I$-band filter should be negligible. 

For each star in the sample, we first re-derive the spectral-type from the strength of the TiO absorption feature at 7000\AA. This is important as RSGs are well-known to be spectrally variable, especially those with later average subtypes. We then determine synthetic photometry for the each of the $I$-band filters relevant for this study: Johnson, Cousins, F814W (HST-WFPC2), and Gunn/Sloan (INT-WFC). 

\begin{table}[htdp]
\caption{The colour correction between different filter systems for an M5-7 supergiant.}
\begin{center}
\begin{tabular}{lc}
\hline \hline
Colour name & Colour (mag) \\
\hline
$I_J - I_C$ & -0.57 \\
$I_J - F814W$ & -0.25 \\
$I_J - {\it Gunn~i^\prime}$ & -0.74 \\
\hline
\end{tabular}
\end{center}
\label{tab:colour}
\end{table}%

In \fig{fig:colours} we plot the results in the form of the difference between Johnson $I$ (hereafter $I_J$) and the various other $I$-bands as a function of spectral type (where 0=M0, 2=M2, -2=K3). We see that these colour corrections are non-negligible, and are a strong function of spectral type. For the purposes of this paper, what we are most interested in is the difference in colour between the various bandpasses for late-type RSGs. To determine these colours, we have used a 2nd-order polynomial to extrapolate the observed trends to a spectral type of M7. We have then defined the average late-type correction to be the average of the extrapolated M7 type and the latest observed type from the M5 star in our sample. These correction factors are listed in Table \ref{tab:colour}.


\bsp	
\label{lastpage}
\end{document}